\documentclass[12pt]{article}
\usepackage{amsmath,amsfonts}
\makeatletter \@addtoreset{equation}{section}

\usepackage{cite}
\usepackage{bm}
\usepackage{dcolumn}
\makeatother
\def\one{{\hbox{ 1\kern-.8mm l}}}
\newcommand{\Dslash}{\not{\hbox{\kern-4pt $D$}}}
\newcommand{\pdslash}{\not{\hbox{\kern-2pt $\partial$}}}
\newcommand{\be}{\begin{equation}}
\newcommand{\bea}{\begin{eqnarray}}
\newcommand{\eea}{\end{eqnarray}}
\newcommand{\ba}{\begin{array}}
\newcommand{\ea}{\end{array}}
\newcommand{\ee}{\end{equation}}

\begin{document}

\begin{titlepage}
\vspace{10mm}
\vspace*{20mm}
\begin{center}
{\Large {\bf On Complexity Growth in Minimal Massive 3D Gravity }\\
}
\vspace*{15mm}
\vspace*{1mm}
{Mohammad M. Qaemmaqami}

\vspace*{1cm}

{ \it  School of Particles and Accelerators\\Institute for Research in Fundamental Sciences (IPM)\\
	P.O. Box 19395-5531, Tehran, Iran }

\vspace*{0.5cm}
{E-mail: m.qaemmaqami@ipm.ir}%

\vspace*{2cm}
%%\maketitle

\vspace*{1cm}

\vspace*{0.5cm}

\vspace*{2cm}
%%\maketitle

\end{center}

\begin{abstract}
We study the complexity growth by using "complexity = action" (CA) proposal in Minimal Massive 3D Gravity(MMG) model which is proposed for resolving the bulk-boundary clash problem of Topologically Massive Gravity(TMG). We observe that the rate of the complexity growth for BTZ black hole saturates the proposed bound by physical mass of the BTZ black hole in the MMG model, when the angular momentum parameter and the inner horizon of black hole goes to zero.

\end{abstract}

\end{titlepage}

\section{Introduction}
One of the holographic conjectures about the inside of black hole is that its growth is dual to the growth of quantum complexity\cite{Susskind:2014rva,Brown:2015bva}. Complexity of a quantum state is defined by the minimum number of simple gates which are needed to build a quantum circuit that constructs them from a reference state.\\

In the context of AdS/CFT duality, One conjecture is "complexity = volume" (CV), the conjecture is an example of  the proposed connection between the tensor network and the geometry of space-time \cite{Swingle:2009bg,Vidal:2007hda,Hartman:2013qma} . The volume of a maximal spacelike slice into the black hole interior, is proportional to the computational complexity of the dual conformal field theory state \cite{Stanford:2014jda},
\begin{eqnarray}
Complexity\sim \frac{V}{G_{N}l_{AdS}},
\end{eqnarray}
where $ V $ is the volume of the Einstein-Rosen bridge(ERB), $ l_{AdS} $ is AdS radius, and $ G_{N} $ is Newton's gravitational constant.\\

The other conjecture is "complexity = action" (CA), the quantum computational complexity of a holographic state is given by the on-shell action on the "Wheeler De-Witt" patch \cite{Brown:2015bva,Brown:2015lvg},
\begin{eqnarray}
\mathcal{C}(\Sigma)= \frac{S_{WDW}}{\pi\hbar},
\end{eqnarray}
where $ \Sigma $ is the time slice which is the intersection of any Cauchy surface in the bulk and the asymptotic boundary. It is conjectured that there is an upper bound on the rate of the complexity growth \cite{Brown:2015lvg}
\begin{eqnarray}
\frac{d\mathcal{C}}{dt}\leq \frac{2M}{\pi\hbar},
\end{eqnarray}
where $ M $ is the mass of black hole. For uncharged black hole the bound is saturated. The above bound is equal to Lloyd's bound which its origin is in quantum computation, if we consider the mass of black hole as the energy of the system.\\
\\
The Aharonov-Anandan-Bohm bound involves the standard deviation of the energy \cite{Anandan:1990fq,Aharonov:1961mka}
\begin{eqnarray}
orthogonality \;\ time\geq \frac{\pi\hbar}{2\Delta E},
\end{eqnarray}

note that the above inequality is similar to Heisenberg's uncertainty of time-energy principle. The above bound is saturated by a two-state quantum system
\begin{eqnarray}
|\psi(t)\rangle =\frac{1}{\sqrt{2}}(|0\rangle+e^{iEt}|E\rangle)\longrightarrow |\langle\psi(t)|\psi(0)\rangle = cos\big(\frac{Et}{2\hbar}\big).
\end{eqnarray}

To generalize and make precise Lloyd's notion of "operations per second", in \cite{Brown:2015lvg} authors considered how complexity builds up in an isolated unitary evolving quantum system in a general quantum state. They proposed a similar bound on the complexity growth rate based on the works of Aharonov-Anandan-Bohm, Margolus-Levitin and Lloyd \cite{Anandan:1990fq,Aharonov:1961mka,Margolus:1997ih,Lloyd:2001bh,Lloyd:2005js,Lloyd:2005dm}. Informally this is the growth rate of the number of simple gates which are needed to construct the state of the computer from a reference state,
\begin{eqnarray}
\frac{d(number\;\ of \;\ gates)}{dt}\leq \frac{2E}{\pi\hbar}.
\end{eqnarray} 

In this paper we study the complexity growth by using "complexity = action" (CA) conjecture in Minimal Massive 3D Gravity(MMG) Model\cite{Bergshoeff:2014pca}, which is proposed for resolving the bulk-boundary clash problem of Topologically Massive Gravity(TMG)\cite{Deser:1981wh}, in some regions of parameter space of the model. The method is adding a new consistent term to the action of the TMG model in the vielbein formalism.\\

In the second section we review shortly the MMG model. In third section we consider the BTZ black hole in MMG model, we introduce the components of vielbein and calculate the non-vanishing components of spin-connection and the auxiliary field of this space-time by solving the equations of motion of MMG model in the vielbein formalism, by plugging the rotating BTZ black hole ansatz. In forth section we consider the well defined variational principle for MMG model and we find the Gibbons-Hawking term for BTZ black hole solution case.\\

In fifth section we calculate the rate of complexity growth by using "complexity = action" (CA) conjecture in MMG model and we observe that the rate of the complexity growth for BTZ black hole saturates the proposed bound by physical mass of the BTZ black hole in the MMG model, when the angular momentum parameter and the inner horizon of black hole goes to zero. Finally we consider the TMG limit of the model and we observe that the rate of the complexity growth saturates  the proposed bound by physical mass of the BTZ black hole in the TMG model, when the inner horizon of black hole goes to zero. The sixth section is devoted to summary.

\section{Minimal Massive 3D Gravity}
The Minimal Massive 3D Gravity(MMG) \cite{Bergshoeff:2014pca} is a proposed model for resolving the bulk-boundary clash problem in the Topologically Massive Gravity(TMG)\cite{Deser:1981wh}, in some region of parameter space of the model. The method is adding a new consistent term to the action of TMG in the vielbein formalism \cite{Bergshoeff:2014pca}. Note that the bulk-boundary clash problem means that; in TMG we have not positive energy of graviton and unitary dual 2D conformal field theory or positive central charges on the boundary at the same time \cite{Bergshoeff:2014pca,Alkac:2017vgg}.
The Lagrangian 3-form of the Minimal Massive 3D Gravity in the vielbein formalism is \cite{Bergshoeff:2014pca}:
\begin{eqnarray}\label{2.1}
L_{MMG}= -\sigma e.R+\frac{\Lambda_{0}}{6}e.e\times e+h.T(\omega)+\frac{1}{2\mu}\big(\omega.d\omega+\frac{1}{3}\omega.\omega\times\omega\big)+\frac{\alpha}{2}e.h\times h,
\end{eqnarray}
where $ e $, is the vielbein, $ \omega $ is the spin-connection and $ h $ is the Lagrange multiplier or the auxiliary field. Note that the dot and the cross mean the internal and the external product respectively, the dot implies contraction of the Lorenz indices of two fields with each other and the cross means contraction of the Lorenz indices of two fields with two indices of the Levi-Civita tensor. And also it is worth noting that the linearized equations of motion of MMG is similar with the linearized equations of motion of TMG by making use a redefinition of topological mass parameter\cite{Tekin:2014jna,Alishahiha:2014dma,Alishahiha:2015whv}. Therefor the model has a single local degree of freedom that is realized as a massive graviton in linearization which is similar to TMG.\\

Note that the action of MMG is only in the vielbein formalism. Because for finding the action in the metric formalism we have to use the equation of motion which is obtained by varying the action with respect to the vielbein field. Therefore we must use the dynamics of the metric for obtaining the action in the metric formalism. Clearly it is not true, because we have to vary the action to find the equations of motion in the metric formalism. Therefore there is the action of MMG only in the vielbein formalism. Then we have to work in the vielbein formalism for calculations in the context of action.\\ 
\\
The equations of motion derived from the action of MMG Eq. (\ref{2.1}) are\cite{Bergshoeff:2014pca}
\bea\label{eq}
T(\omega) - \alpha e\times h =0,\;\;\;\;\;\;\;\;\;\; R(\omega)+\mu e\times h +\sigma\mu T(\omega) = 0,
\eea
and
\bea\label{eq1}
-\sigma R(\omega)+\frac{\Lambda_{0}}{2} e\times e +D(\omega)h+\frac{\alpha}{2} h\times h =0.
\eea
The curvature and the torsion 2-forms are defined as follows:
\begin{eqnarray}
R(\omega)=d\omega+\frac{1}{2}\omega\times\omega,   \;\;\;\;\;\;\      \;\;\;\;\;\;\    T(\omega)=de+\omega\times e.
\end{eqnarray}

To proceed we note that although for generic $ \alpha\neq 0 $ the torsion, $ T(\omega) $ is non-zero,
one may define a new torsion free spin-connection, $ \Omega=\omega+\alpha h $ by which the Lagrangian
3-form reads \cite{Alishahiha:2015whv}:
\begin{eqnarray}\label{Eq2.5}
L_{MMG}(e,\Omega, h) &=& -\sigma e\cdot R(\Omega) +\frac{\Lambda_{0}}{6} 
e\cdot e\times e + h\cdot T(\Omega) +\frac{1}{2\mu}
(\Omega\cdot d \Omega+\frac{1}{3}\Omega\cdot \Omega \times \Omega)-\cr
&&
-\frac{1}{2\mu} 
\bigg(2\alpha h\cdot R(\Omega)
-\alpha^2 h\cdot Dh+\frac{\alpha^3}{3}h\cdot h\times h\bigg)+
\cr &&+\sigma\alpha e\cdot Dh
-\frac{\alpha}{2}(1+\sigma\alpha)e\cdot h\times h.
\end{eqnarray}

In this notation, assuming to have a well defined variation principle, by varying the above action with respect to the three fields, $ (e, \Omega, h) $ we obtain the corresponding equations of motion as follows,
\begin{eqnarray}\label{Eq2.6}
&&T(\Omega)=0,   \;\;\;\;\     \;\;\;\;\   R(\Omega)+\frac{\alpha\Lambda_{0}}{2}e\times e+\mu(1+\sigma\alpha)^{2}e\times h=0, \cr\nonumber\\
&&D(\Omega)h-\frac{\alpha}{2}h\times h+\sigma\mu(1+\sigma\alpha)e\times h+\frac{\Lambda_{0}}{2}e\times e=0.
\end{eqnarray}

Here the covariant derivative is defined by $ D(\Omega)A=dA+\Omega\times A $, where the spin-connection, $ \Omega $ is obtained from first equation of motion, $ T(\Omega)=de+\Omega\times e= 0 $, or in other words the torsion-free condition.

\section{BTZ Black hole}

The metric of the rotating BTZ black hole is given by \cite{Banados:1992wn},
\begin{eqnarray}
ds^{2}=-\frac{(r^{2}-r_{+}^{2})(r^{2}-r_{-}^{2})}{l^{2}r^{2}}dt^{2}+\frac{l^{2}r^{2}}{(r^{2}-r_{+}^{2})(r^{2}-r_{-}^{2})}dr^{2}+r^{2}\bigg(d\varphi-\frac{r_{+}r_{-}}{lr^{2}}dt\bigg)^{2}.
\end{eqnarray}

The mass and the angular momentum parameter of the BTZ black hole, in terms of inner and outer horizons of the black hole and AdS radius, respectively are:
\begin{eqnarray}
M= \frac{r_{+}^{2}+r_{-}^{2}}{8Gl^{2}},  \;\;\;\;\;\;\      \;\;\;\;\;\;\   J= \frac{r_{+}r_{-}}{4Gl}.
\end{eqnarray}
The non-vanishing components of the vielbein of the rotating BTZ black hole are:
\begin{eqnarray}\label{3.3}
&&e_{t}~^{0}=\frac{\sqrt{(r^{2}-r_{+}^{2})(r^{2}-r_{-}^{2})}}{lr},  \;\;\;\;\;\;\        \;\;\;\;\;\;\   e_{t}~^{2}=-\frac{r_{+}r_{-}}{lr},\cr\nonumber\\
&&e_{r}~^{1}= \frac{lr}{\sqrt{(r^{2}-r_{+}^{2})(r^{2}-r_{-}^{2})}},  \;\;\;\;\;\;\        \;\;\;\;\;\;\   e_{\varphi}~^{2}=r.
\end{eqnarray}

Note that the relation between the metric and the vielbein field is; $ g_{\mu\nu}= \eta_{ab}e_{\mu}~^{a}e_{\nu}~^{b} $, where $ \eta_{ab} $, is the local Minkowski metric\\

By the first equation of motion, Eq. (\ref{Eq2.6}) which is the torsion free condition, one can find the non-vanishing components of the spin-connection from the vielbein field Eq. (\ref{3.3}) as follows,
\begin{eqnarray}
&&\Omega_{t}~^{2}=-\frac{r}{l},  \;\;\;\;\;\;\        \;\;\;\;\;\;\   \Omega_{r}~^{1}   =-\frac{r_{+}r_{-}}{r\sqrt{(r^{2}-r_{+}^{2})(r^{2}-r_{-}^{2})}},\cr\nonumber\\
&&\Omega_{\varphi}~^{0}=-\frac{\sqrt{(r^{2}-r_{+}^{2})(r^{2}-r_{-}^{2})}}{rl},   \;\;\;\;\;\;\        \;\;\;\;\;\;\   \Omega_{\varphi}~^{2}= \frac{r_{+}r_{-}}{lr}.
\end{eqnarray}

In following we can find the auxiliary field "$ h_{\mu}~^{a} $", from the second equation of motion, Eq. (\ref{Eq2.6}), we can rewrite the equation in this form:
\begin{eqnarray}
R(\Omega)+e\times h^{'}=0, \;\;\;\;\;\;\     \;\;\;\;\;\;\    h^{'}= \frac{\alpha\Lambda_{0}}{2}e+\mu(1+\sigma\alpha)^{2}h,
\end{eqnarray}
from the above equation we can find the "$ h^{'}_{\mu}~^{a} $" \cite{Merbis:2014vja},
\begin{eqnarray}
h^{'}_{\mu}~^{a}=-det(e)^{-1}\epsilon^{\lambda\rho\sigma}(e_{\lambda}~^{a}e_{\mu b}-\frac{1}{2}e_{\mu}~^{a}e_{\lambda b})R_{\rho\sigma}~^{b},
\end{eqnarray}
therefore one can find auxiliary field "$ h_{\mu}~^{a} $" as follows
\begin{eqnarray}\label{3.7}
h_{\mu}~^{a}= -\frac{1}{\mu(1+\sigma\alpha)^{2}}\bigg[det(e)^{-1}\epsilon^{\lambda\rho\sigma}(e_{\lambda}~^{a}e_{\mu b}-\frac{1}{2}e_{\mu}~^{a}e_{\lambda b})R_{\rho\sigma}~^{b}+\frac{\alpha\Lambda_{0}}{2}e_{\mu}~^{a}\bigg].
\end{eqnarray}
We find the non vanishing components of  $ h_{\mu}~^{a} $, by replacing the vielbein and spin-connection fields in the above relation Eq. (\ref{3.7}), as follows:
\begin{eqnarray}
&&h_{t}~^{0}=\frac{1-\alpha l^{2}\Lambda_{0}}{2\mu l^{2}(1+\sigma\alpha)^{2}}\frac{\sqrt{(r^{2}-r_{+}^{2})(r^{2}-r_{-}^{2})}}{lr},    \;\;\;\;\;\;\        \;\;\;\;\;\;\  h_{t}~^{2}=-\frac{1-\alpha l^{2}\Lambda_{0}}{2\mu l^{2}(1+\sigma\alpha)^{2}}\frac{r_{+}r_{-}}{lr},\cr\nonumber\\
&&h_{r}~^{1}= \frac{1-\alpha l^{2}\Lambda_{0}}{2\mu l^{2}(1+\sigma\alpha)^{2}}\frac{lr}{\sqrt{(r^{2}-r_{+}^{2})(r^{2}-r_{-}^{2})}},  \;\;\;\;\;\;\        \;\;\;\;\;\;\   h_{\varphi}~^{2}=\frac{1-\alpha l^{2}\Lambda_{0}}{2\mu l^{2}(1+\sigma\alpha)^{2}}r.
\end{eqnarray}
One can see easily for the BTZ black hole we have,
\begin{eqnarray}\label{3.9}
h_{\mu}~^{a}=\bigg(\frac{1-\alpha l^{2}\Lambda_{0}}{2\mu l^{2}(1+\sigma\alpha)^{2}}\bigg)e_{\mu}~^{a}.
\end{eqnarray}

We can use this relation for finding the Gibbons-Hawking term in BTZ black hole case, by a well defined variational principle in Dirichlet boundary condition.
\section{Variational Principle}
The equations of motion, Eq. (\ref{Eq2.6}) rely on the fact that
the model admits a well-imposed variational principle. This procedure requires the proper Gibbons-Hawking term to make sure that all boundary terms can be consistently removed. In this section we would like to reexamine the well defined variation of the action leading to the corresponding equations of motion.\\

To proceed let us consider the action of the MMG model Eq (\ref{Eq2.5}) whose variation with respect to the fields $ e $, $ \Omega $ and $ h $ are given by\cite{Alishahiha:2015whv};
\begin{eqnarray}
&& \delta_{e} L(e,\Omega, h) =E_{e}\cdot \delta e-D(\Omega)(h\cdot \delta e),\cr
&&\delta_{\Omega} L(e,\Omega, h) = E_{\Omega}\cdot \delta\Omega 
+D(\Omega)\left(\sigma e\cdot \delta\Omega-
\frac{1}{2\mu}(\Omega\cdot \delta\Omega-2\alpha h\cdot \delta\Omega)\right),
\cr
&&\delta_{h} L(e,\Omega, h)= E_{h}\cdot \delta h+
D(\Omega)\left(-\sigma\alpha e\cdot \delta h -\frac{1}{2\mu}
(\alpha^2 h\cdot \delta h -2\alpha \Omega\cdot \delta h)\right).
\end{eqnarray} 
 
Using the Stokes' theorem the corresponding  boundary terms appearing in the above variation may be recast to the following form
\begin{eqnarray}\label{4.2}
\delta S|_{boundary} &=& \frac{1}{8\pi G}\int_{\partial\mathcal{M}}d^{2}x
\epsilon^
{ij}\bigg(
h_{i a}
\delta e_{j}^{a}+\big[\sigma e-\frac{1}{2\mu}\Omega
+\frac{\alpha}{\mu}h\big]_{i a}\delta\Omega_{j}^{a}+\cr
&&~+\big[-\alpha\sigma e+\frac{\alpha}{\mu}\Omega
-\frac{\alpha^{2}}{2\mu}h\big]_{i a}\delta h_{j}^{a}\bigg).
\end{eqnarray}

For a well defined variational principle, we need a Gibbons-Hawking term to cancel the boundary terms in the Dirichlet boundary condition. In general finding the Gibbons-Hawking term is difficult for this model, but in the case of BTZ black hole solution, the "$ h_{\mu}~^{a} $" field is proportional with the vielbein field, "$ e_{\mu}~^{a} $", Eq. (\ref{3.9}) therefore the variation of the auxiliary field "$ h_{\mu}~^{a} $" is as follows:
\begin{eqnarray}
\delta h_{\mu}~^{a}=\bigg(\frac{1-\alpha l^{2}\Lambda_{0}}{2\mu l^{2}(1+\sigma\alpha)^{2}}\bigg)\delta e_{\mu}~^{a},
\end{eqnarray}
then the third term of variation of action on the boundary Eq. (\ref{4.2}) is canceled by Dirichlet boundary condition therefore the Gibbons-Hawking term in this case is:
\begin{eqnarray}\label{4.4}
S_{GH}&=& -\frac{1}{8\pi G}\int_{\partial\mathcal{M}} d^{2}x
\epsilon^{ij}\big(\sigma\tilde{e}+\frac{\alpha}{\mu}\tilde{h}\big)_{i a}\tilde{\Omega}_{j}^{a}=\cr\nonumber\\ &=&-\frac{1}{8\pi G}\bigg[\sigma+\alpha\bigg(\frac{1-\alpha l^{2}\Lambda_{0}}{2\mu^{2} l^{2}(1+\sigma\alpha)^{2}}\bigg)\bigg]\int_{\partial\mathcal{M}} d^{2}x
\epsilon^{ij}\tilde{e}_{i a}\tilde{\Omega}_{j}^{a},
\end{eqnarray}
where $ \tilde{e} $, $ \tilde{\Omega} $ and $ \tilde{h} $ are the boundary vielbein, the boundary spin-connection and the boundary auxiliary field, respectively.

\section{Complexity Growth in Minimal Massive 3D Gravity}

In the proposal known as "complexity = action" (CA) the quantum computational complexity of a holographic state is given by the on-shell action evaluated on a bulk region known as the "Wheeler-De Witt" patch \cite{Brown:2015bva,Brown:2015lvg}
\begin{eqnarray}\label{5.1}
\mathcal{C}(\Sigma)= \frac{S_{WDW}}{\pi\hbar},
\end{eqnarray}
it is conjectured that there is an upper bound on the rate of the complexity growth \cite{Brown:2015lvg}
\begin{eqnarray}
\frac{d\mathcal{C}}{dt}\leq \frac{2M}{\pi\hbar},
\end{eqnarray}
where $ M $ is the mass of black hole. For uncharged black hole the bound is saturated. Note that here we use "complexity = action" (CA) conjecture, because the bound on complexity growth rate in "complexity = action" (CA) is exactly the Lloyd's bound. And also it is worth noting that, however the first proposal was "complexity = volume" (CV), but community have paid more attention to "complexity = action" (CA) conjecture and people have done more works in this context.\\

The corresponding action of Minimal Massive 3D Gravity containing the Gibbons-Hawking term is given by
\begin{eqnarray}
S=S_{\mathcal{M}}+S_{\partial\mathcal{M}}&=&\frac{1}{16\pi G}\int_{\mathcal{M}}d^{3}x \epsilon^{\lambda\mu\nu}[L_{MMG}(e,\Omega, h)]_{\lambda\mu\nu}-\cr
&&-\frac{1}{8\pi G}\bigg[\sigma+\alpha\bigg(\frac{1-\alpha l^{2}\Lambda_{0}}{2\mu^{2} l^{2}(1+\sigma\alpha)^{2}}\bigg)\bigg]\int_{\partial\mathcal{M}} d^{2}x
\epsilon^{ij}\tilde{e}_{i a}\tilde{\Omega}_{j}^{a},
\end{eqnarray}
where $ [L_{MMG}(e,\Omega, h)]_{\lambda\mu\nu} $ is the Lagrangian 3-form of MMG which is defined in second section by Eq. (\ref{Eq2.5}).\\

To compute the rate of the complexity growth, we should calculate the difference between the on-shell actions which are evaluated over two nearby WDW patches \cite{Alishahiha:2017hwg}. In the present case by BTZ solution at the late time, only the region between the inner and outer horizons contributes to this difference. Therefore we find;
\begin{eqnarray}
&&\delta S_{\mathcal{M}}= S_{\mathcal{M}}\big[WDW|_{t+\delta t}\big]- S_{\mathcal{M}}\big[WDW|_{t}\big]=\cr\nonumber\\
&&=\frac{1}{16\pi G}\int_{t}^{t+\delta t}\int_{r_{-}}^{r^{+}}\int_{0}^{2\pi} \epsilon^{\lambda\mu\nu}[L_{MMG}(e,\Omega, h)]_{\lambda\mu\nu}dtdrd\varphi=\cr\nonumber\\
&&=-\frac{2\pi\delta t}{16\pi G}\bigg[\sigma+\alpha\bigg(\frac{1-\alpha l^{2}\Lambda_{0}}{2\mu^{2} l^{2}(1+\sigma\alpha)^{2}}\bigg)\bigg]\int_{r_{-}}^{r^{+}}\frac{4r}{l^{2}}=-\bigg[\sigma+\alpha\bigg(\frac{1-\alpha l^{2}\Lambda_{0}}{2\mu^{2} l^{2}(1+\sigma\alpha)^{2}}\bigg)\bigg]\bigg(\frac{r_{+}^{2}-r_{-}^{2}}{4Gl^{2}}\bigg)\delta t.\cr\nonumber\\
\end{eqnarray}
In following the contribution of the Gibbons-Hawking term Eq. (\ref{4.4}) is given by
\begin{eqnarray}
\delta S_{\partial\mathcal{M}}&=&-\frac{1}{8\pi G}\bigg[\sigma+\alpha\bigg(\frac{1-\alpha l^{2}\Lambda_{0}}{2\mu^{2} l^{2}(1+\sigma\alpha)^{2}}\bigg)\bigg]\int_{\partial\mathcal{M}} d^{2}x
\epsilon^{ij}\tilde{e}_{i a}\tilde{\Omega}_{j}^{a}=\cr\nonumber\\
&=&\frac{1}{8\pi G}\bigg[\sigma+\alpha\bigg(\frac{1-\alpha l^{2}\Lambda_{0}}{2\mu^{2} l^{2}(1+\sigma\alpha)^{2}}\bigg)\bigg]\bigg(\int_{t}^{t+\delta t}\int_{0}^{2\pi}\frac{1}{l^{2}}\big(2r^{2}-r_{+}^{2}-r_{-}^{2}\big)dtd\varphi|_{r_{+}}-\cr\nonumber\\
&&-\int_{t}^{t+\delta t}\int_{0}^{2\pi}\frac{1}{l^{2}}\big(2r^{2}-r_{+}^{2}-r_{-}^{2}\big)dtd\varphi|_{r_{-}}\bigg)=\cr\nonumber\\
&&=\bigg[\sigma+\alpha\bigg(\frac{1-\alpha l^{2}\Lambda_{0}}{2\mu^{2} l^{2}(1+\sigma\alpha)^{2}}\bigg)\bigg]\bigg(\frac{r_{+}^{2}-r_{-}^{2}}{2Gl^{2}}\bigg)\delta t.
\end{eqnarray}
Therefore, we find the rate of the complexity growth by Eq. (\ref{5.1})
\begin{eqnarray}
\dot{\mathcal{C}}=\frac{1}{\pi\hbar}\frac{dS}{dt}\equiv \frac{1}{\pi\hbar}\bigg(\frac{dS_{\mathcal{M}}}{dt}+\frac{dS_{\partial\mathcal{M}}}{dt}\bigg)=\frac{1}{\pi\hbar}\bigg[\sigma+\alpha\bigg(\frac{1-\alpha l^{2}\Lambda_{0}}{2\mu^{2} l^{2}(1+\sigma\alpha)^{2}}\bigg)\bigg]\frac{r_{+}^{2}-r_{-}^{2}}{4Gl^{2}}.
\end{eqnarray}
We know the mass of BTZ black hole in MMG theory\cite{Nam:2016pfp,Setare:2015vea,Yekta:2015gja}
\begin{eqnarray}
\mathcal{M}_{MMG}= \bigg[\sigma+\alpha\bigg(\frac{1-\alpha l^{2}\Lambda_{0}}{2\mu^{2} l^{2}(1+\sigma\alpha)^{2}}\bigg)\bigg]\bigg(\frac{r_{+}^{2}+r_{-}^{2}}{8Gl^{2}}\bigg)+\frac{r_{+}r_{-}}{4G\mu l^{3}},
\end{eqnarray}
one can see for $ r_{-}=0 $, which is the non rotating BTZ black hole case in 3D Einstein Gravity (the angular momentum parameter is zero), we have
\begin{eqnarray}
\dot{\mathcal{C}}=\frac{1}{\pi\hbar}\frac{dS}{dt}=\frac{2\mathcal{M}_{MMG}}{{\pi\hbar}},
\end{eqnarray}
it's interesting, one can see the rate of the complexity growth saturates the proposed bound \cite{Brown:2015lvg} by the physical mass of the BTZ black hole in the Minimal Massive 3D Gravity model.\\
\\
Finally let's consider the TMG limit of the model when $ \alpha=0 $,
\begin{eqnarray}\label{5.9}
\dot{\mathcal{C}}= \frac{1}{\pi\hbar}\frac{dS}{dt}\equiv \frac{1}{\pi\hbar}\bigg(\frac{dS_{\mathcal{M}}}{dt}+\frac{dS_{\partial\mathcal{M}}}{dt}\bigg)=\frac{\sigma}{\pi\hbar}\bigg(\frac{r_{+}^{2}-r_{-}^{2}}{4Gl^{2}}\bigg).
\end{eqnarray}
We know the mass mass of BTZ black hole in TMG theory \cite{Bouchareb:2007yx},\cite{Nam:2016pfp},
\begin{eqnarray}
\mathcal{M}_{TMG}=\sigma\bigg(\frac{r_{+}^{2}+r_{-}^{2}}{8Gl^{2}}\bigg)+\frac{r_{+}r_{-}}{4G\mu l^{3}},
\end{eqnarray}
one can see for $ r_{-}=0 $, we have
\begin{eqnarray}\label{5.11}
\dot{\mathcal{C}}=\frac{1}{\pi\hbar}\frac{dS}{dt}=\frac{2\mathcal{M}_{TMG}}{\pi\hbar},
\end{eqnarray}
clearly the rate of the complexity growth saturates the proposed bound \cite{Brown:2015lvg} by the physical mass of the BTZ black hole in the Topologically Massive Gravity model.

\section{Summary}
In this work we study the complexity growth by using "complexity = action" (CA) conjecture in Minimal Massive 3D Gravity(MMG) Model which is proposed for resolving the bulk-boundary clash problem of Topologically Massive Gravity(TMG). We observe that the rate of the complexity growth for BTZ black hole saturates the proposed bound by physical mass of the BTZ black hole in the MMG model, when the angular momentum parameter and the inner horizon of black hole goes to zero. We can say it is another evidence for the hypothesis that black holes are the fastest computers in the nature\cite{Brown:2015bva}, as they are the fastest scramblers\cite{Sekino:2008he}.\\

Recently some work have been done on quantum complexity in context of holography and black holes \cite{Alishahiha:2015rta,Chapman:2016hwi,Alishahiha:2017hwg,Cai:2016xho,Pan:2016ecg,Yang:2016awy,Carmi:2016wjl,Kim:2017lrw,Cai:2017sjv,Bakhshaei:2017qud,Abad:2017cgl,Reynolds:2017lwq,Tao:2017fsy,Guo:2017rul,Alishahiha:2017cuk,Nagasaki:2017kqe,Miao:2017quj,Ghodrati:2017roz}, and also it needs more investigations and calculations in the other Gravity models as higher derivative Gravity models and more black hole solutions, in addition it seems interesting to study the connection between holographic complexity and tensor networks or butterfly effect in different context and models to achieve deeper understanding of these important and interesting phenomenon of the nature.\\

\textbf{Note added}: As this work was being completed, \cite{Ghodrati:2017roz} appeared which has minor overlap with present paper.

\section*{Acknowledgment}
I am grateful to thank Ali Naseh for useful discussions. I also thank the referee for her/his useful comments. I would like to thank the Institute for Research in Fundamental Sciences(IPM) and the Science Ministry of IRAN for research facilities and financial support.

%Bibliography

%--------------------------------------------------------------------
\end{document}